\documentstyle[prl,aps,axodraw,psfig,floats]{revtex}

\def\bi{\bibitem}

\def\ni{\noindent}
\def\beb{\begin{thebibliography}{999}}
\def\eeb{\end{thebibliography}}
\def\bei{\begin{itemize}}
\def\eei{\end{itemize}}
\def\bef{\begin{figure}}
\def\eef{\end{figure}}
\def\ben{\begin{enumerate}}
\def\een{\end{enumerate}}
\def\beq{\begin{equation}}
\def\eeq{\end{equation}}
\def\ber{\begin{eqnarray}}
\def\eer{\end{eqnarray}}
\def\twidle{\widetilde}
\def\f{\frac}
\begin{document}
\draft
\twocolumn[
\hsize\textwidth\columnwidth\hsize\csname@twocolumnfalse\endcsname
\title{Absorption of Electro-magnetic Waves in a Magnetized Medium}
\author{Avijit K.~Ganguly$^1$, Sushan Konar$^2$}
\address{$^1$ Saha Institute of Nuclear Physics, 1/AF, Bidhan-Nagar, Calcutta 700064, India \\
$^2$IUCAA, Post Bag 4, Ganeshkhind, Pune 411007, India\\
e-mail : avijit@tnp.saha.ernet.in, sushan@iucaa.ernet.in}
\maketitle
%%%%%%%%%%%%%%%%%%%%%%%%%
\begin{abstract} 
In continuation to our earlier work, in which the structure of the vacuum polarisation
tensor in a medium was analysed in presence of a background electro-magnetic field, we discuss
the absorptive part of the vacuum polarization tensor. Using the real time formalism of finite 
temperature field theory we calculate the absorptive part of 1-loop vacuum polarisation tensor 
in the weak field limit ($eB < m^2$). Estimates of the absorption probability are also made for 
different physical conditions of the background medium.
\end{abstract}
%%%%%%%%%%%%%%%%%%%%%%%%%
%\pacs{PACS numbers::~04.70Dy, 12.20.-m, 04.62.+v}
\narrowtext
\bigskip
]
%%%%%%%%%%%%%%%%%%%%%%%%%
\section{Introduction}
\label{intr}
%%%%%%%%%%%%%%%%%%%%%%%%%

\ni Processes in a magnetised plasma are of interest in widely different physical systems ranging from
laboratory plasma to astrophysical objects~\cite{melr}. Therefore, the propagation of electro-magnetic
waves, in a magnetised plasma, continues to evoke much interest. Fortunately, almost all of the terrestrial
or the astrophysical systems, barring the newly discovered `magnetars'~\cite{kouv}, have magnetic fields
smaller than the QED limit ($eB < m_e^2$ i.e, $B \le 10^{13}$~Gauss). This allows for a weak-field treatment
of the plasma processes relevant to such physical situations. Moreover, this treatment is also valid for 
compact astrophysical objects (viz. white dwarfs or neutron stars) for which the Landau level spacings are
quite small compared to the electron Fermi energy~\cite{chan}. This ensures that the magnetic field does not
introduce any spatial anisotropy in the collective plasma behaviour. \\

\ni In view of this, in an earlier paper~\cite{GKP1} (paper-I henceforth) we analysed the structure of the 
vacuum polarization tensor (denoted by $\Pi_{\mu \nu}$) in a background medium in presence of a uniform
external magnetic field, in conformity with Lorenz and gauge invariance. In the weak-field limit we retained
terms up-to ${\cal O(B)}$ to obtain the field dependence of the vacuum polarisation, calculated at the 1-loop
level. As expected, we recovered Faraday rotation, i.e, the phenomenon of the rotation of the plane of 
polarisation of an electro-magnetic wave passing through an ionised medium in presence of an external magnetic
field, from the ${\cal O(B)}$ term. \\

\ni The present paper can be thought of as a sequel to paper-I, in which, we calculated the dispersive part 
of $\Pi_{\mu \nu}$. In this work we calculate the absorptive part thereby obtaining a complete expression 
for the polarisation at the 1-loop level. Once again we retain terms up-to ${\cal O(B)}$ to investigate the 
field dependence of the absorptive part which can be further used to evaluate the damping/instability of the 
photons propagating in a plasma. It is worth noting that the damping/instability probability of a propagating 
photon due to pair creation/annihilation in a medium can also be obtained from the tree-level diagram using 
the exact wave-functions of the fermions in an external magnetic field. However, these wave-functions are not 
well defined in the weak-field limit and therefore use of tree-level amplitudes may not be very accurate. On 
the other hand the general structure of $\Pi_{\mu \nu}$ has a well defined limit as $e{\cal B} \rightarrow 0$. 
Hence this approach is free from the problems of the weak-field limit. \\

\ni The organization of the document is as follows. In section-\ref{form} we discuss the basic formalism. 
Section-\ref{calc} contains the details of the calculation of the absorptive part of the polarisation tensor in 
presence of a background magnetic field. In section-\ref{disp} we outline the relation between the polarisation 
tensor and the dispersion relation. Finally in section-\ref{limit} we discuss our results to a few limiting
cases. The appendix contains a few details.

%%%%%%%%%%%%%%%%%%%%%%%%%
\section{Formalism}
\label{form}
%%%%%%%%%%%%%%%%%%%%%%%%%

\ni Recall that in presence of an external field, the interaction part of the Lagrangian is given by:
\beq
{\cal{L}}= - \, \int j^{\mu}(x) \, {\cal{A}}_{\mu}(x) d^4x \,,
\eeq
where $j^{\mu}(x)$ can be defined in terms of the fermion field $\Psi(x)$ (solutions of the 
equation of motion in presence of an external field) as:
\beq
j^{\mu}(x) = - \, e\bar{\Psi}(x) \gamma^{\mu}\Psi(x) \,. 
\eeq
Correspondingly, the $S$-matrix for the theory is defined to be:
\beq
S = T e^{i\int d^4x \, j^{\mu}(x) \, {\cal{A}}_{\mu}(x)} \,,
\eeq
where $T$ refers to the time-ordering of the product. Further, the $S$-matrix can be written in the 
following form:
\beq
S = 1+\sum_{n \ge 1} \frac{i^n}{n!} \int d^4x_1 ..d^4x_n \, T({\cal{L}}(x_1)..{\cal{L}}(x_n)) \,.
\eeq
It is worth noting that the second term in the right hand side is the usual ${\cal{T}}$-matrix. 
Since we are interested in fermionic pair-creation/annihilation in presence of a magnetic field in a
medium, terms up to $O(e^2)$ are retained, $e$ being the coupling constant. Expanding the 
$\cal{T}$-matrix up to second order in the coupling constant, we obtain:
\beq
{\cal{T}}^{2} - {\cal{T}}^{2\dagger} = i {\cal{T}}^{1}{\cal{T}}^{1\dagger} \,,
\label{unit2}
\eeq
using the unitarity of the $S$-matrix. Here the superscripts on $\cal{T}$ denote the order of the
coupling constant in the expansion. Taking the expectation value of eq.(\ref{unit2}) between two 
in-states we have:
\beq
\sum_{out}| <out| {\cal{T}}^{1}|in> |^{2} = 2 Im <in|{\cal{T}}^{2} |in> \,.
\label{unit3}
\eeq
Depending on the choice of the initial states, we obtain the decay/production amplitude of the 
particles.\\

\ni In order to find these rates in a thermal medium the thermal expectation value of eq.(\ref{unit3}) 
has to be evaluated. There are a number of ways to perform the thermal averaging. Amongst them the more 
commonly used formalisms are the imaginary time technique of Matsubara \cite{matsu} and the real time 
finite temperature formalism (see \cite{larry,kapusta,lebellac,adas,kobes,jose,land} and references 
therein). In the present work we use the real time formalism of the finite temperature field theory. The 
propagator acquires a matrix structure in this formalism and the off-diagonal elements provide the 
decay/production amplitudes. Unfortunately, the direct evaluation of the off-diagonal elements in presence 
of an external field is rather complicated. Therefore, for the ease of calculation, we work with the
11-component of the propagator and find imaginary part of the 11-component of the photon polarisation
tensor ($\Pi^{11}_{\mu \nu}$). This quantity, multiplied by appropriate factors then gives the correct
value for the imaginary part of the polarisation tensor~\cite{adas,kobes,jose,land}. Though for 
notational brevity we shall suppress the 11-superscript for both the propagator and the polarisation 
tensor in the rest of the paper. \\

%%%%%%%%%%%%%%%%%%%%%%%%%%%%%%
\bef
\begin{center}
\begin{picture}(150,50)(0,-25)
\Photon(0,0)(40,0){2}{4}
\Text(20,5)[b]{$k\rightarrow$}
\Photon(110,0)(150,0){2}{4}
\Text(130,5)[b]{$k\rightarrow$}
\Text(75,30)[b]{$p+k\equiv p'$}
\Text(75,-30)[t]{$p$}
\SetWidth{1.2}
\Oval(75,0)(25,35)(0)
\ArrowLine(74,25)(76,25)
\ArrowLine(76,-25)(74,-25)
\end{picture}
\end{center}
\caption[]{One-loop diagram for the vacuum polarization.}\label{f:1loop}
\eef
%%%%%%%%%%%%%%%%%%%%%%%%%

\ni At the 1-loop level, the vacuum polarization tensor arises from the diagram in fig.~\ref{f:1loop}. 
The dominant contribution to the vacuum polarization comes from the electron line in the loop. To 
evaluate this diagram we use the electron propagator within a thermal medium in presence of a background 
electro-magnetic field. Rather than working with a completely general 
background field we specialize to the case of a purely magnetic field. Once this is assumed, the 
field can be taken in the $z$-direction without any further loss of generality. We denote the 
magnitude of this field by $\cal B$. Ignoring at first the presence of the medium, the electron 
propagator in such a field can be written down following Schwinger's approach~\cite{Schwing,Tsai,Dittrich}:
\ber
i S_B^V(p) &=& \int_0^\infty ds\, {e^{ie{\cal B}s\sigma\!_z} \over \cos(e{\cal B}s)} \nonumber \\
&\times& \exp \left[ is \left( p_\parallel^2 - {\tan (e{\cal B}s) \over e{\cal B}s} 
\, p_\perp^2 - m^2 + i\epsilon \right) \right] \nonumber \\
&\times& \left( \rlap/p_\parallel - \frac{e^{-ie{\cal B}s\sigma_z}} {\cos(e{\cal B}s)}\rlap/ p_\perp + m \right) \,,
\label{SV}
\eer
where 
\ber
\rlap/ p_\parallel &=& \gamma_0 p_0 - \gamma_3 p_3 \\
p_\parallel^2 &=& p_0^2 - p_3^2 \\
\rlap/p_\perp &=& \gamma_1 p_1 + \gamma_2 p_2 \\
p_\perp^2 &=& p_1^2 + p_2^2 \,,
\eer
and $\sigma_z$ is given by:
\beq
\sigma_z = i\gamma_1 \gamma_2 = - \gamma_0 \gamma_3 \gamma_5 \,,
\label{sigz} 
\eeq
where the two forms are equivalent because of the definition of $\gamma_5$.
Since
\beq
e^{ie{\cal B}s\sigma_z} = \cos \; e{\cal B}s + i\sigma_z \sin \; e{\cal B}s \,,
\eeq
we can rewrite the propagator in the following form: 
\beq
i S_B^V(p) = \int_0^\infty ds\; e^{\Phi(p,s)} C(p,s) \,,
\label{SV2}
\eeq
where we have used the shorthands,
\ber
\Phi(p,s) &\equiv& is \left( p_\parallel^2 - {\tan (e{\cal B}s) \over e{\cal B}s} 
\, p_\perp^2 - m^2 \right) - \epsilon |s| \,, 
\label{Phi} \\
C(p,s) &\equiv& \Big[ ( 1 + i\sigma_z \tan  e{\cal B}s ) (\rlap/p_\parallel + m ) 
- (\sec^2 e{\cal B}s) \rlap/ p_\perp \Big] \,. 
\label{C}
\eer
Of course in the range of integration indicated in eq.(\ref{SV2}) $s$ is never negative and hence $|s|$ equals $s$. 
It should be mentioned here that we follow the notation adopted in paper-I to ensure continuity. In the presence of 
a background medium, the above propagator is modified to~\cite{Elmf}:
\beq
iS(p) = iS_B^V(p) - \eta_F(p) \left[ iS_B^V(p) - i\overline S_B^V(p) \right] \,,
\label{fullprop}
\eeq
where 
\beq
\overline S_B^V(p) \equiv \gamma_0 S^{V \dagger}_B(p) \gamma_0 \,,
\label{Sbar}
\eeq
for a fermion propagator and $\eta_F(p)$ contains the distribution function for the fermions and the anti-fermions:
\ber
\eta_F(p) &=& \Theta(p\cdot u) f_F(p,\mu,\beta) \nonumber \\
&+& \Theta(-p\cdot u) f_F(-p,-\mu,\beta) \,.  
\label{eta} 
\eer
Here, $f_F$ denotes the Fermi-Dirac distribution function:
\beq
f_F(p,\mu,\beta) = {1\over e^{\beta(p\cdot u - \mu)} + 1} \,,
\eeq
and $\Theta$ is the step function given by:
\ber
\Theta(x) &=& 1, \; \mbox{for $x > 0$} \,, \nonumber \\
&=& 0, \; \mbox{for $x < 0$} \,. \nonumber
\eer
Rewriting eq.(\ref{fullprop}) in the following form:
\ber
iS(p) &=& \frac{i}{2} \left[ S_B^V(p) + \overline S_B^V(p) \right] \nonumber \\
&& + i (1/2 - \eta_F(p)) \left[ S_B^V(p) - \overline S_B^V(p) \right] \,, \nonumber \\
&=& iS_{\rm re} + iS_{\rm im}
\label{S_reim}
\eer
we recognise:
\ber
S_{\rm re} &=&  \frac{1}{2} \left[ S_B^V(p) + \overline S_B^V(p) \right] \, ,\\
S_{\rm im} &=& (1/2 - \eta_F(p)) \left[ S_B^V(p) - \overline S_B^V(p) \right] \, ;
\eer
where the subscripts {\em re} and {\em im} refer to the real and imaginary parts of the propagator. 
Using the form of $S_B^V(p)$ in eq.(\ref{SV2}) we obtain the imaginary part to be:
\ber
iS_{\rm im} &=& (1/2 - \eta_F(p)) \left[ iS_B^V(p) - i\overline S_B^V(p) \right] \, , \nonumber \\
&=& (1/2 - \eta_F(p)) \int_{-\infty}^\infty ds\; e^{\Phi(p,s)} C(p,s) \,.
\label{Sim}
\eer
with $\Phi(p,s)$ and $C(p,s)$ defined by eq.s.(\ref{Phi}) and (\ref{C}). \\

%%%%%%%%%%%%%%%%%%%%%%%%%%%%%%%%%%%%%%%%%%%%%%%%%%%%%%%%%%%%%%%%%%%%%
\section{Calculation of the 1-loop vacuum polarization}
\label{calc}
%%%%%%%%%%%%%%%%%%%%%%%%%%%%%%%%%%%%%%%%%%%%%%%%%%%%%%%%%%%%%%%%%%%%%

%%%%%%%%%%%%%%%%%%%%%%%%%%%%%%%%%%%%%%%%%%%%%%%%%%%%%%%%%%%%%%%%%%%%%
\subsection{Identifying the Relevant Terms}
%%%%%%%%%%%%%%%%%%%%%%%%%%%%%%%%%%%%%%%%%%%%%%%%%%%%%%%%%%%%%%%%%%%%%

\ni The amplitude of the 1-loop diagram of fig.~\ref{f:1loop} can be written as:
\beq
i \Pi_{\mu\nu}(k) = - \int \frac{d^4p}{(2\pi)^4} (ie)^2 \; \mbox{tr}\, 
\left[\gamma_\mu \, iS(p) \gamma_\nu \, iS(p')\right] \,,
\eeq
where, for the sake of notational simplicity, we have used
\beq
p' = p+k \,.
\label{p'}
\eeq
The minus sign on the right side is for a closed fermion loop and $S(p)$ is the 
propagator given by eq.(\ref{fullprop}). This implies: 
\beq
\Pi_{\mu\nu}(k) = -ie^2 \int \frac{d^4p}{(2\pi)^4} \; 
\mbox{tr}\, \left[\gamma_\mu \, iS(p) \gamma_\nu \, iS(p')\right] \,.
\label{1loopampl}
\eeq
Using eq.(\ref{S_reim}) we have:
\ber
\Pi_{\mu\nu}(k) &=& -ie^2 \int \frac{d^4p}{(2\pi)^4} \nonumber \\
&\times& \mbox{tr}\, [\gamma_\mu \, (iS_{\rm re}(p) + iS_{\rm im}(p)) \nonumber \\
&& \times \, \gamma_\nu \, (iS_{\rm re}(p') + iS_{\rm im}(p') ] \,.
\eer
Then the absorptive part of the polarisation tensor is given by:
\ber
\Pi_{\mu\nu}^{11}(k) &=& -ie^2 \int \frac{d^4p}{(2\pi)^4} \; \mbox{tr}\, 
\left[\gamma_\mu \, iS_{\rm im}(p) \gamma_\nu \, iS_{\rm im}(p^{\prime}) \right] \\
&=& -ie^2 \int \frac{d^4p}{(2\pi)^4} \; (1/2 - \eta_F(p)) \, (1/2 - \eta_F(p^{\prime})) \nonumber \\
&\times& \int_{-\infty}^\infty ds\; e^{\Phi(p,s)} \,
\int_{-\infty}^\infty ds^{\prime}\; e^{\Phi(p^{\prime},s^{\prime})} \nonumber \\
&\times& \mbox{tr}\, \left[\gamma_\mu \, C(p,s) \gamma_\nu \, C(p^{\prime},s^{\prime}) \right] \nonumber \\
&=& - \frac{ie^2}{4} \, \int \frac{d^4p}{(2\pi)^4} \, X(\beta, k, p.u) \nonumber\\
&\times&\int_{-\infty}^\infty ds\; e^{\Phi(p,s)} \, \int_{-\infty}^\infty ds^{\prime}
\; e^{\Phi(p^{\prime},s^{\prime})} \nonumber\\
&\times& \mbox{tr}\, \left[\gamma_\mu \, C(p,s) \gamma_\nu \, C(p^{\prime},s^{\prime}) \right] 
\label{PiT}
\eer
where we have defined:
\beq
X(\beta, k, p) = (1 - 2 \eta_F(p)) \, (1 - 2 \eta_F(p^{\prime})) \,.
\label{X}
\eeq

%%%%%%%%%%%%%%%%%%%%%%%%%%%%%%%%%%%%%%%%%%%%%%%%%%%%%%%%%%%%%%%%%%%%%%
\subsection{Extracting the Gauge Invariant Piece}
%%%%%%%%%%%%%%%%%%%%%%%%%%%%%%%%%%%%%%%%%%%%%%%%%%%%%%%%%%%%%%%%%%%%%%

\ni As discussed earlier we are interested only in terms up-to ${\cal O(B)}$ and therefore shall drop all terms of
higher order from the subsequent calculations. Now, notice that the phase factors appearing in eq.(\ref{PiT}) are 
even in $\cal B$. Thus, to keep terms up to ${\cal O(B)}$ we need consider only the odd terms from the traces. Performing 
the traces the odd terms come out to be:
\ber
\Pi^{11}_{\mu\nu}(k, \beta) &=& - \, ie^2 \, \int \frac{d^4p}{(2\pi)^4} X(\beta, k, p) \int_{-\infty}^\infty ds e^{\Phi(p,s)} 
\nonumber \\
&\times& \int_{-\infty}^\infty ds' \; e^{\Phi(p',s')} R_{\mu\nu} \,;
\eer
where
\ber
R_{\mu\nu} &=& \varepsilon_{\mu\nu03} m^2 
\big( \tan e{\cal B}s - \tan e{\cal B}s' \big) \nonumber\\*
&& + \, \varepsilon_{\mu\nu\alpha_\parallel\beta_\parallel} 
\Big( p^{\widetilde\alpha_\parallel} p'^{\beta_\parallel} 
\tan e{\cal B}s - p'^{\widetilde\alpha_\parallel} 
p^{\beta_\parallel} \tan e{\cal B}s' \Big) \nonumber\\*
&& + \, \varepsilon_{\mu\nu\alpha_\parallel\beta_\perp} 
\Big( p^{\widetilde\alpha_\parallel} p'^{\beta_\perp} 
\tan e{\cal B}s \sec^2 e{\cal B}s' \nonumber \\
&& - \, p'^{\widetilde\alpha_\parallel} p^{\beta_\perp} \tan e{\cal B}s' \sec^2 e{\cal B}s \Big) \,.
\eer
In writing this expression, we have used the notation $p^{\widetilde\alpha_\parallel}$, for example. 
This signifies a component of $p$ which can take only the `parallel' indices, i.e., 0 and 3, and is 
moreover different from the index $\alpha$ appearing elsewhere in the expression. Using now the 
definition of $p'$ from eq. (\ref{p'}), we can write,
\beq
R_{\mu\nu} = R^{(1)}_{\mu\nu} + R^{(2)}_{\mu\nu} \,,
\eeq
where,
\beq
R^{(1)}_{\mu\nu} = \varepsilon_{\mu\nu\alpha_\parallel\beta} 
\Big[ p^{\widetilde\alpha_\parallel} \tan e{\cal B}s 
+ p'^{\widetilde\alpha_\parallel} \tan e{\cal B}s' \Big] k^\beta 
\label{R1}
\eeq
and,
\ber
R^{(2)}_{\mu\nu} &=& \tan e{\cal B}s \Bigg[ m^2 \varepsilon_{\mu\nu03} 
+ \varepsilon_{\mu\nu\alpha_\parallel\beta_\parallel} p^{\widetilde\alpha_\parallel} p^{\beta_\parallel} \nonumber \\
&+& \varepsilon_{\mu\nu\alpha_\parallel\beta_\perp} \Big( p^{\widetilde\alpha_\parallel} p^{\beta_\perp} 
+ p^{\widetilde\alpha_\parallel} p'^{\beta_\perp} \tan^2 e{\cal B}s' \Big) \Bigg] \nonumber\\* 
&-& \tan e{\cal B}s' \Bigg[ m^2 \varepsilon_{\mu\nu03} 
+ \varepsilon_{\mu\nu\alpha_\parallel\beta_\parallel} p'^{\widetilde\alpha_\parallel} p'^{\beta_\parallel} \nonumber \\
&+& \varepsilon_{\mu\nu\alpha_\parallel\beta_\perp} \Big( p'^{\widetilde\alpha_\parallel} p'^{\beta_\perp} 
+ p'^{\widetilde\alpha_\parallel} p^{\beta_\perp} \tan^2 e{\cal B}s' \Big) \Bigg] \,. 
\label{R2}
\eer
Obviously, $R^{(1)}_{\mu\nu}$ is gauge invariant, i.e., $k^\mu R^{(1)}_{\mu\nu}=k^\nu R^{(1)}_{\mu\nu}=0$. To simplify 
the other term, we first note that the combinations in which the parallel components of $p$ and $p'$ appear in 
eq.(\ref{R2}) can be simplified by using the following identity:
\ber
\varepsilon_{\mu\nu\alpha_\parallel\beta_\parallel} a^{\widetilde\alpha_\parallel} b^{\beta_\parallel} 
= - \, \varepsilon_{\mu\nu03} \; a \cdot b_\parallel \,,
\eer
which holds for any two vectors $a$ and $b$. Moreover, we perform the calculations in the rest frame of 
the medium where $p\cdot u=p_0$. Thus the distribution function does not depend on the spatial components 
of $p$. In the last term of each square bracket of eq.(\ref{R2}), the integral over the transverse 
components of $p$ has the following generic structure:
\beq
\int d^2 p_\perp \; e^{\Phi(p,s)} e^{\Phi(p',s')} \times 
\mbox{($p^{\beta_\perp}$ or $p'^{\beta_\perp}$)} \,.
\eeq
Notice that,
\ber
{\partial \over \partial p_{\beta_\perp}} \Big[ \; e^{\Phi(p,s)} e^{\Phi(p',s')} \Big] 
&=& \Big( \tan e{\cal B}s \; p^{\beta_\perp} + \tan e{\cal B}s' \; p'^{\beta_\perp} \Big) \nonumber \\
&\times& {2i\over e{\cal B}} e^{\Phi(p,s)} e^{\Phi(p',s')} \,.
\label{single_derivative}
\eer
However, this expression, being a total derivative, should integrate to zero. Thus we obtain that,
\beq
\tan e{\cal B}s \; p^{\beta_\perp} \stackrel\circ= - \tan e{\cal B}s' \; p'^{\beta_\perp} \,,
\eeq
where the sign `$\stackrel\circ=$' means that the expressions on both sides of it, though not necessarily 
equal algebraically, yield the same integral. This gives,
\ber
p^{\beta_\perp} &\stackrel\circ=& - \, {\tan e{\cal B}s' \over \tan e{\cal B}s 
+ \tan e{\cal B}s'} \; k^{\beta_\perp} \,,\nonumber\\*
p'^{\beta_\perp} &\stackrel\circ=&  {\tan e{\cal B}s \over \tan e{\cal B}s 
+ \tan e{\cal B}s'} \; k^{\beta_\perp} \,.
\eer
Using these identities we rewrite eq. (\ref{R2}) in the following form:
\beq
R^{(2)}_{\mu\nu} = R^{(2a)}_{\mu\nu} + \varepsilon_{\mu\nu03} R^{(2b)} \,,
\eeq
where,
\ber
R^{(2a)}_{\mu\nu} &=& - \varepsilon_{\mu\nu\alpha_\parallel\beta}
\; {\tan e{\cal B}s \; \tan e{\cal B}s' \over \tan e{\cal B}(s+s')} 
\; (p+p')^{\widetilde\alpha_\parallel} k^\beta \,,\\
R^{(2b)} &=& (m^2 - p_\parallel^2) \tan e{\cal B}s - (m^2 - p_\parallel^{\prime2}) \tan e{\cal B}s' \nonumber \\
&-& {\tan e{\cal B}s \; \tan e{\cal B}s' \over \tan e{\cal B}(s+s')} \; (p+p') \cdot k_\parallel \,.
\eer
The term called $R^{(2b)}$ does not vanish on contraction with arbitrary $k^\mu$. This term is not gauge 
invariant, and therefore must vanish on integration. It can be shown, following the arguments in paper-I, 
that this term indeed vanishes. Therefore the contribution to the absorptive part of the vacuum polarization 
tensor which is odd in $\cal B$ is given by:
\ber
\Pi^{11}_{\mu\nu}(k, \beta) &=& - \, ie^2 \, \int \frac{d^4p}{(2\pi)^4} X(\beta, k, p)
\int_{-\infty}^\infty ds \; e^{\Phi(p,s)} \nonumber \\
&\times& \int_{-\infty}^\infty ds' \; e^{\Phi(p',s')} \Big[ R^{(1)}_{\mu\nu} + R^{(2a)}_{\mu\nu} \Big] \nonumber\\
&=& - \, ie^2 \varepsilon_{\mu\nu\alpha_\parallel\beta} k^\beta \, \int \frac{d^4p}{(2\pi)^4} X(\beta, k, p) \nonumber \\
&\times& \int_{-\infty}^\infty ds \; e^{\Phi(p,s)} \int_{-\infty}^\infty ds' \; e^{\Phi(p',s')} \nonumber\\*
&\times& \Bigg[ p^{\widetilde\alpha_\parallel} \tan e{\cal B}s 
+ p'^{\widetilde\alpha_\parallel} \tan e{\cal B}s' \nonumber \\
&-& {\tan e{\cal B}s \; \tan e{\cal B}s' \over \tan e{\cal B}(s+s')} \; (p+p')^{\widetilde\alpha_\parallel} \Bigg] \,.
\label{oddpart}
\eer
%

%%%%%%%%%%%%%%%%%%%%%%%%%%%%%%%%%%%%%%%%%%%%%%%%%%%%%%%%%%%%%%%%%
\subsection{Terms Linear in External Field }
%%%%%%%%%%%%%%%%%%%%%%%%%%%%%%%%%%%%%%%%%%%%%%%%%%%%%%%%%%%%%%%%%
\ni Eq.(\ref{oddpart}) is one of the important results obtained in this paper. This part of the polarisation 
tensor is odd in powers of the external field and is purely matter dominated. Retaining terms only up-to 
${\cal O(B)}$ in eq.(\ref{oddpart}) we arrive at:
\ber
\Pi^{11}_{\mu\nu}(k, \beta) &=& - \, ie^3 {\cal B} \varepsilon_{\mu\nu\alpha_\parallel\beta} k^\beta \, 
\int \frac{d^4p}{(2\pi)^4} X(\beta, k, p) \nonumber \\
&\times& \int_{-\infty}^\infty ds \; e^{\Phi(p,s)} \int_{-\infty}^\infty ds' \; e^{\Phi(p',s')} \nonumber \\
&\times& \Bigg[ p^{\widetilde\alpha_\parallel} \frac{s^2}{s+s'} 
+ p'^{\widetilde\alpha_\parallel} \frac{s'^2}{s+s'} \Bigg] \,.
\eer
Making the transformation $p \rightarrow -p-k$ in the first piece we obtain:
\ber
\Pi^{11}_{\mu\nu}(k, \beta) &=& - \, e^3 {\cal B} \varepsilon_{\mu\nu\alpha_\parallel\beta} k^\beta 
\int \frac{d^4p}{(2\pi)^4} Y(\beta, k, p_0) \nonumber \\
&\times& \partial_{k^{\widetilde\alpha_\parallel}} I(p,p')\,,
\eer
where,
\beq
I(p,p') = \int_{-\infty}^\infty ds \; e^{\Phi(p,s)} \int_{-\infty}^\infty ds' \; e^{\Phi(p',s')} \frac{s'}{s+s'} \,,
\label{I}
\eeq
and
\beq
Y(\beta, k, p_0) = X(\beta, k, p) - X(\beta, k, -p') \,.
\label{Y}
\eeq
Notice that $p$ appears in $X(\beta, k, p)$ only as a combination $p.u$. Therefore, only $p_0$ is present in $X$ for a 
medium in its rest frame. We use this fact to define $Y(\beta, k, p_0)$. From the form of $Y(\beta, k, p_0)$ it is evident
that $\Pi^{11}_{\mu\nu}(k, \beta)$ to ${\cal O(B)}$ vanishes in absence of a medium. Since the behaviour of the dispersive
part of the polarisation tensor is similar~\cite{GKP1} we conclude that the term linearly dependent on the external magnetic
field does not contribute to the total polarisation tensor in absence of a medium. \\

\ni It is shown in the Appendix that $I(p,p')$ can be reduced to the following form:
\beq
I(p,p') = 2 \pi^2 \delta(p^2 -m^2)  \delta(p'^2 -m^2) \,.
\eeq
Therefore, we finally have:
\ber
\Pi^{11}_{\mu\nu}(k, \beta) 
&=& - \, 2 \pi^2 e^3 {\cal B} \varepsilon_{\mu\nu\alpha_\parallel\beta} k^\beta \, \int \frac{d^4p}{(2\pi)^4} Y(\beta, k, p_0) 
\nonumber \\
&\times& \delta(p^2 - m^2) \, \partial_{k^{\widetilde\alpha_\parallel}} \delta(p'^2 - m^2) \,.
\label{Pfinal}
\eer
The delta-functions can be written in the following form:
\ber
\delta(p^2 - m^2) &=& \frac{1}{2 E_p} [\delta(p_0 - E_p) + \delta(p_0 + E_p)] \,, \nonumber \\
\delta(p'^2 - m^2) &=& \delta(p^2_0 + k_0^2 + 2 p_0 k_0 - E_{p'}^2) \,, 
\label{delta}
\eer
with the definitions, 
\ber
E^2_p &=& m^2 + P^2 \,, \nonumber \\
E^2_{p'} &=& m^2 + P^2 + K^2 + 2 P K \cos \theta \,,
\eer
where $P$ and $K$ are the magnitudes of the spatial parts of $p$ and $k$, $\theta$ being the angle between $\vec P$ 
and $\vec K$. Using eq.s(\ref{delta}) to perform the $dp_0$ and $d\phi$ integral we obtain:
\ber
\Pi^{11}_{\mu\nu}(k, \beta) &=& - \, \pi^2 e^3 {\cal B} \varepsilon_{\mu\nu\alpha_\parallel\beta} k^\beta 
\int \frac{P^2 dP \, d(\cos \theta)}{(2\pi)^2} \frac{1}{E_p} \nonumber \\
&\times& \Big[ Y(\beta, k, E_p) \, \partial_{k^{\widetilde\alpha_\parallel}} \delta(k^2 + 2 k_0 E_p - 2 P K \cos \theta) 
\nonumber \\
&+& (E_p \rightarrow - E_p) \Big]\\
\label{pi_gen1}
&=& - \, \frac{\pi^2}{2} e^3 {\cal B} \varepsilon_{\mu\nu\alpha_\parallel\beta} k^\beta 
\, \partial_{k^{\widetilde\alpha_\parallel}} \, \frac{1}{K} \int \frac{P dP 
\, d(\cos \theta)}{(2\pi)^2} \frac{1}{E_p} \nonumber \\
&\times& \Big[\delta(\cos \theta - \frac{k^2 + 2 k_0 E_p}{2 P K}) \, Y(\beta, k, E_p) \nonumber \\
&+& (E_p \rightarrow - E_p) \Big] \nonumber \\
&+& \frac{\pi^2}{2} e^3 {\cal B} \varepsilon_{\mu\nu\alpha_\parallel\beta} \frac{k^\beta}{K}
\int \frac{P dP \, d(\cos \theta)}{(2\pi)^2} \frac{1}{E_p} \nonumber \\
&\times& \Big[\delta(\cos \theta - \frac{k^2 + 2 k_0 E_p}{2 P K}) 
\, \partial_{k^{\widetilde\alpha_\parallel}} Y(\beta, k, E_p) \nonumber \\
&+& (E_p \rightarrow - E_p) \Big] \,.
\label{pi_gen2}
\eer
Now performing the $\theta$ integration we arrive at:
\ber
\Pi^{11}_{\mu\nu}(k, \beta)
&=& - \, \frac{\pi^2}{2} e^3 {\cal B} \varepsilon_{\mu\nu\alpha_\parallel\beta} k^\beta 
\, \partial_{k^{\widetilde\alpha_\parallel}} \,\frac{1}{K} \int \frac{P dP}{2\pi} \frac{1}{E_p} \nonumber \\
&\times& \left\{Y(\beta, k, E_p) + Y(\beta, k, - E_p)\right\} \nonumber \\
&+& \frac{\pi^2}{2} e^3 {\cal B} \varepsilon_{\mu\nu\alpha_\parallel\beta} 
\frac{k^\beta}{K} \int \frac{P dP}{2\pi} \frac{1}{E_p} \nonumber \\
&\times& \partial_{k^{\widetilde\alpha_\parallel}} \left\{Y(\beta, k, E_p) + Y(\beta, k, - E_p)\right\} \,.
\label{e_pi-2}
\eer
It should be noted that eq.(\ref{pi_gen2}) is valid only if the following conditions are met with. Since, 
inside the delta function $-1 \le cos \theta \le 1$ as $0 \le \theta \le \pi$, the range of integration 
over $P$ must contain the interval $[P_{\rm min}, P_{\rm max}]$, where,
\beq
P_{\rm min} = - \frac{K}{2} + \frac{k_0}{2} (1- \frac{4 m^2}{k^2})^{1/2}  
\label{pmin}
\eeq
\beq
P_{\rm max} = \frac{K}{2} + \frac{k_0}{2} (1- \frac{4 m^2}{k^2})^{1/2} \,. 
\label{pmax}
\eeq
Since $P$ is real the condition $k^2 \ge 4m^2$ must be satisfied. {\it This is an important kinematic 
constraint which ensures the conservation of energy-momentum in the weak field limit.} In the next sections 
we shall discuss a few specific background media to find the modification of the absorptive part of the 
polarisation tensor in presence of a background magnetic field. \\

\ni It has long been realised that the amplitudes computed using the imaginary-time formalism, when continued 
analytically to real frequencies, usually differ from those computed using the 11-component of the real-time 
technique~\cite{kobes}. In order to make the results consistent with each other we need to multiply our result 
by $\tanh \beta k_0/2$ to obtain:
\ber
\Pi^{\rm abs}_{\mu\nu}(k, \beta)
&=& - \, \frac{\pi^2}{2} e^3 {\cal B} \varepsilon_{\mu\nu\alpha_\parallel\beta} k^\beta \tanh \frac{\beta k_0}{2} \nonumber \\
&\times& \partial_{k^{\widetilde\alpha_\parallel}} \,\frac{1}{K} \int \frac{P dP}{2\pi} \frac{1}{E_p} \nonumber \\
&\times& \left\{Y(\beta, k, E_p) + Y(\beta, k, - E_p)\right\} \nonumber \\
&+& \frac{\pi^2}{2} e^3 {\cal B} \varepsilon_{\mu\nu\alpha_\parallel\beta} \frac{k^\beta}{K} \tanh \frac{\beta k_0}{2} 
\int \frac{P dP}{2\pi} \frac{1}{E_p} \nonumber \\
&\times& \partial_{k^{\widetilde\alpha_\parallel}} \left\{Y(\beta, k, E_p) + Y(\beta, k, - E_p)\right\} \,.
\label{pi_final}
\eer

\ni Now eq.(\ref{pi_final}) can be further simplified using Leibnitz theorem for differentiation of an integral given by:
\ber
\frac{d}{dc} \int^{b(c)}_{a(c)} F(x,c)dx &=& \int^{b(c)}_{a(c)} \frac{\partial}{\partial c} F(x,c)dx \nonumber \\
&& + \, F(b,c)\frac{db}{dc} - F(a,c) \frac{da}{dc} \,.
\label{Leb}
\eer
Using eq.s(\ref{pmin}) and (\ref{pmax}) we can reduce eq.(\ref{pi_final}) to the following form:
\ber
&& \Pi^{\rm abs}_{\mu\nu}(k, \beta) \nonumber \\
&=& - \, \frac{\pi^2}{2} e^3 {\cal B} \varepsilon_{\mu\nu\alpha_\parallel\beta} k^\beta \, \tanh \frac{\beta k_0}{2}
\, \left(\partial_{k^{\widetilde\alpha_\parallel}} \, \frac{1}{K}\right) \nonumber \\
&& \times \, \int_{P_{\rm min}}^{P_{\rm max}} \frac{P dP}{2\pi} \frac{1}{E_p} 
\, \left\{Y(\beta, k, E_p) + Y(\beta, k, - E_p)\right\} \nonumber \\
&& - \,  \frac{\pi^2}{2} e^3 {\cal B} \varepsilon_{\mu\nu\alpha_\parallel\beta} \frac{k^\beta}{K} \, \tanh \frac{\beta k_0}{2} \nonumber \\
&& \times \, \left[\frac{P}{E_p} \left\{Y(\beta, k, E_p) + Y(\beta, k, - E_p)\right\}\right]_{P_{\rm max}} 
\, \partial_{k^{\widetilde\alpha_\parallel}} P_{\rm max} \nonumber \\
&& + \, \frac{\pi^2}{2} e^3 {\cal B} \varepsilon_{\mu\nu\alpha_\parallel\beta} \frac{k^\beta}{K} \, \tanh \frac{\beta k_0}{2} \nonumber \\
&& \times \, \left[\frac{P}{E_p} \, \left\{Y(\beta, k, E_p) + Y(\beta, k, - E_p)\right\}\right]_{P_{\rm min}} 
\, \partial_{k^{\widetilde\alpha_\parallel}} P_{\rm min}. \nonumber \\
\label{e_pi-1}
\eer
One should note the difference in sign between the different terms of the polarisation tensor in eq.(\ref{e_pi-1}).  
This opens up the possibility that for a given magnetic field, depending on the value of chemical potential and 
the external photon momentum, $\Pi^{\rm abs}_{\mu\nu}(k, \beta) $ can be either positive or negative giving rise to
damping or instability of the propagating photon.
%
%%%%%%%%%%%%%%%%%%%%%%%%%%%%%%%%%%%%%%%%%%%%%%%%%%%%%%%%%%%%%%%%%%%%%
\section{Dispersion Relation}
\label{disp}
%%%%%%%%%%%%%%%%%%%%%%%%%%%%%%%%%%%%%%%%%%%%%%%%%%%%%%%%%%%%%%%%%%%%%%

\ni In this section we recapitulate the dispersion relations for a photon traveling in a magnetized medium (see paper-I for details). 
The quadratic part of the Lagrangian with quantum and thermal correction is --
\beq
{\cal L} = {1\over 2} \left[ - k^2 \widetilde g_{\mu\nu} + \Pi_{\lambda\rho} (k) \right]  A^\mu (k) A^\nu(-k) \,,
\label{L0+Pi}
\eeq
where $\widetilde g_{\mu\nu} = \left( g_{\mu \nu} - \frac{k_\mu k_\nu}{k^2} \right)$. In momentum space the equation of motion takes the 
form:
\beq
\left[ (-k^2+\omega_0^2) g_{\mu \nu} + \Pi^{tot}_{\mu\nu} \right] A^\nu = 0 \,.
\label{eqmotion}
\eeq
Here we assume the Lorenz gauge condition ($\partial_\mu A^{\mu} = 0$) and $\omega_0$ is the plasma frequency defined by:
\beq
\omega_0^2 = 4 e^2 \int \frac{d^3p}{(2\pi)^3 2 E_p} \left(1 - \frac{P^2}{3E_p^2}\right) [f_+ + f_-] \,,
\eeq
where we have introduced the notation \beq
f_\pm = f_F(|p_0|, \mp \mu)\,.
\eeq
The quantity $\Pi^{tot}_{\mu\nu}$ is the total vacuum polaristaion tensor containing the dispersive as well as the 
absorptive part. Assuming the direction of photon propagation to be along the direction of the magnetic field, for 
the two transverse components of the photon field $A^{\mu}$, we obtain the following dispersion relations:
\beq
k^2 = \omega_0^2 \pm \left( a_{\rm disp} + a_{\rm abs} \right) \,,
\label{disp1}
\eeq
where $a_{\rm disp}$ and $a_{\rm abs}$ are the dispersive and the absorptive parts of $\Pi_{12}$ respectively. 
Recall that a circularly polarised wave undergoes an exponential damping through the factor $e^{-{\rm Im} k_0 t}$. 
Therefore, the lifetime $\tau$ of a photon is given by the imaginary part of $k_0$ where $k_0$ is given by:
\beq
k^2_0 = \omega_0^2+ K^2 \pm  a_{\rm disp} \mp i a_{\rm abs}\,. 
\label{disp2}
\eeq
%
%%%%%%%%%%%%%%%%%%%%%%%%%%%%%%%%%%%%%%%%%%%%%%%%%%%%%%%%%%%%%%%%%%%%%%
\section{Limiting Cases}
\label{limit}
%%%%%%%%%%%%%%%%%%%%%%%%%%%%%%%%%%%%%%%%%%%%%%%%%%%%%%%%%%%%%%%%%%%%%%
%%%%%%%%%%%%%%%%%%%%%%%%%%%%%%%%%%%%%%%%%%%%%%%%%%%%%%%%%%%%%%%%%%%%%%
\subsection{Non-Relativistic Fermion--Anti-Fermion Plasma}
%%%%%%%%%%%%%%%%%%%%%%%%%%%%%%%%%%%%%%%%%%%%%%%%%%%%%%%%%%%%%%%%%%%%%%

\ni Let us consider the case of a non-degenerate fermion--anti-fermion plasma. Using eq.(\ref{eta}) we can write:
\beq
2\eta_F(p) - 1 = \epsilon (p.u) \tanh \left(\frac{\beta (p.u - \mu)}{2} \right) \,,
\eeq
where,
\ber
\epsilon(x) &=& 1, \; \mbox{for $x > 0$} \,, \nonumber \\
&=& - 1, \; \mbox{for $x < 0$} \,. 
\eer
This renders the thermal factors in the polarisation tensor in the following form:
\ber
&&(1 - 2 \eta_F(p)) \, (1 - 2 \eta_F(p^{\prime})) = \epsilon (p.u) \, \epsilon (p^{\prime}.u) \nonumber \\
&&\times \Bigg[ 1 - \coth \frac{\beta k.u}{2} \, 
\Big(\tanh \frac{\beta (p^{\prime}.u - \mu)}{2} \nonumber \\
&&\;\;\;\;\;\;\;\;\;\;\;\;\;\;\;\;\;\;\;\;\;\;\;\;\;\;\;\; - \tanh \frac{\beta (p.u - \mu)}{2} \Big) \Bigg] \,, 
\label{thermal}
\eer
where we have used the following identity:
\ber
1 - \tanh A \tanh B = \coth (A-B) \, (\tanh A - \tanh B) \,. \nonumber
\eer
Notice that in eq.(\ref{thermal}) only the second term within the square bracket contributes to the finite 
temperature effect, the first one giving the zero temperature case. Using eq.s (\ref{X}), (\ref{Y}), (\ref{thermal}) and using the relation
$\tanh z = 1 - \frac{2}{e^{2z}+1}$ one can write,
\ber
Y(\beta, k, p_0) &=& 2 \, \epsilon (p_0) \, \epsilon(p'_0) \coth \frac{\beta k_0}{2} \nonumber \\ 
&\times& \Bigg[\frac{1}{e^{\beta(p_0 + \mu)}+1} - \frac{1}{e^{\beta(p_0 - \mu)} + 1} \nonumber \\
&& + \, \frac{1}{e^{\beta(p'_0 - \mu)}+1} - \frac{1}{e^{\beta(p'_0 + \mu)} + 1} \Bigg] \,.
\label{Y3}
\eer

\ni
Therefore, for the thermal part, using eq. (\ref{Y3}) we obtain,
\ber
&& Y(\beta, k, E_p) + Y(\beta, k, -E_p) \nonumber \\
&=& 2 \, \epsilon(E_p) \, \coth \frac{\beta k_0}{2} \nonumber \\
&\times& \Bigg(\epsilon(E_p+ k_0) \big(\frac{1}{e^{\beta(E_p +k_0 - \mu)}+1} - \frac{1}{e^{\beta(E_p +k_0 + \mu)}+1} \nonumber \\
&& - \frac{1}{e^{\beta(E_p - \mu)}+1} + \frac{1}{e^{\beta(E_p + \mu)}+1} \big) \nonumber \\
&+& \epsilon(E_p- k_0) \big(\frac{1}{e^{\beta(E_p -k_0 - \mu)}+1} - \frac{1}{e^{\beta(E_p -k_0 + \mu)}+1} \nonumber \\
&& - \frac{1}{e^{\beta(E_p - \mu)}+1} + \frac{1}{e^{\beta(E_p + \mu)}+1} \big) \Bigg)
\label{Y_f}
\eer
\ni
For a non-relativistic non-degenerate plasma one can approximate $E_p$ inside
the integration by the particle mass $m$. For a highly energetic photon with $k_0 >> m$ we can write:   
\ber
&& Y(\beta, k, E_p) + Y(\beta, k, -E_p)
=
2 coth\f{\beta k_0}{2} \nonumber \\ &\times&
\Bigg( 
\bigg[ \frac{1}{e^{\beta(m+k_0 -\mu)}+1}
-\frac{1}{e^{\beta(m+k_0+\mu)}+1}
\nonumber \\
&-&  
 \frac{1}{e^{\beta(m -k_0 -\mu)}+1}
+\frac{1}{e^{\beta(m - k_0+\mu)}+1}
\bigg]
\Bigg) \nonumber \\
&&
\simeq 2 coth\f{\beta k_0}{2} 
\bigg( e^{ \beta k_0} - e^{- \beta k_0}\bigg)\nonumber \\ &\times&
\bigg( e^{ -\beta(m+\mu)} - e^{- \beta(m -\mu)}\bigg). 
\label{ndg-pls}
\eer
Using the form of the thermal factors given by eq.~(\ref{ndg-pls})
in eq.~(\ref{e_pi-1}) the absorptive part of the polarisation
tensor becomes:
\ber
\Pi^{\rm abs}_{\mu\nu}(k, \beta) 
 &=& - \,\pi^2 e^3 {\cal B} \varepsilon_{\mu\nu\alpha_\parallel\beta} k^\beta \,
\bigg( e^{ \beta k_0} - e^{- \beta k_0}\bigg)
\nonumber \\ &\times&
\bigg( e^{ -\beta(m+\mu)} - e^{- \beta(m -\mu)}\bigg)
\,
\nonumber \\ &\times&
 \Bigg[
\left(\partial_{k^{\widetilde\alpha_\parallel}} \, \frac{1}{K}\right) 
(\frac{K}{2 \pi})
 - \partial_{k^{\widetilde\alpha_\parallel}}K
\Bigg]
\label{ngd-absp}
\eer
From eq.~(\ref{ngd-absp}) one can easily check that in the limit of $m << \mu$
the absorptive part of the polarisation tensor becomes negative signifying damping of the propagating photon. 
%%%%%%%%%%%%%%%%%%%%%%%%%%%%%%%%%%%%%%%%%%%%%%%%%%%%%%%%%%%%%%%%%%%%%%
\subsection{Degenerate Fermion Gas}
%%%%%%%%%%%%%%%%%%%%%%%%%%%%%%%%%%%%%%%%%%%%%%y%%%%%%%%%%%%%%%%%%%%%%%%

\bef
\psfig{file=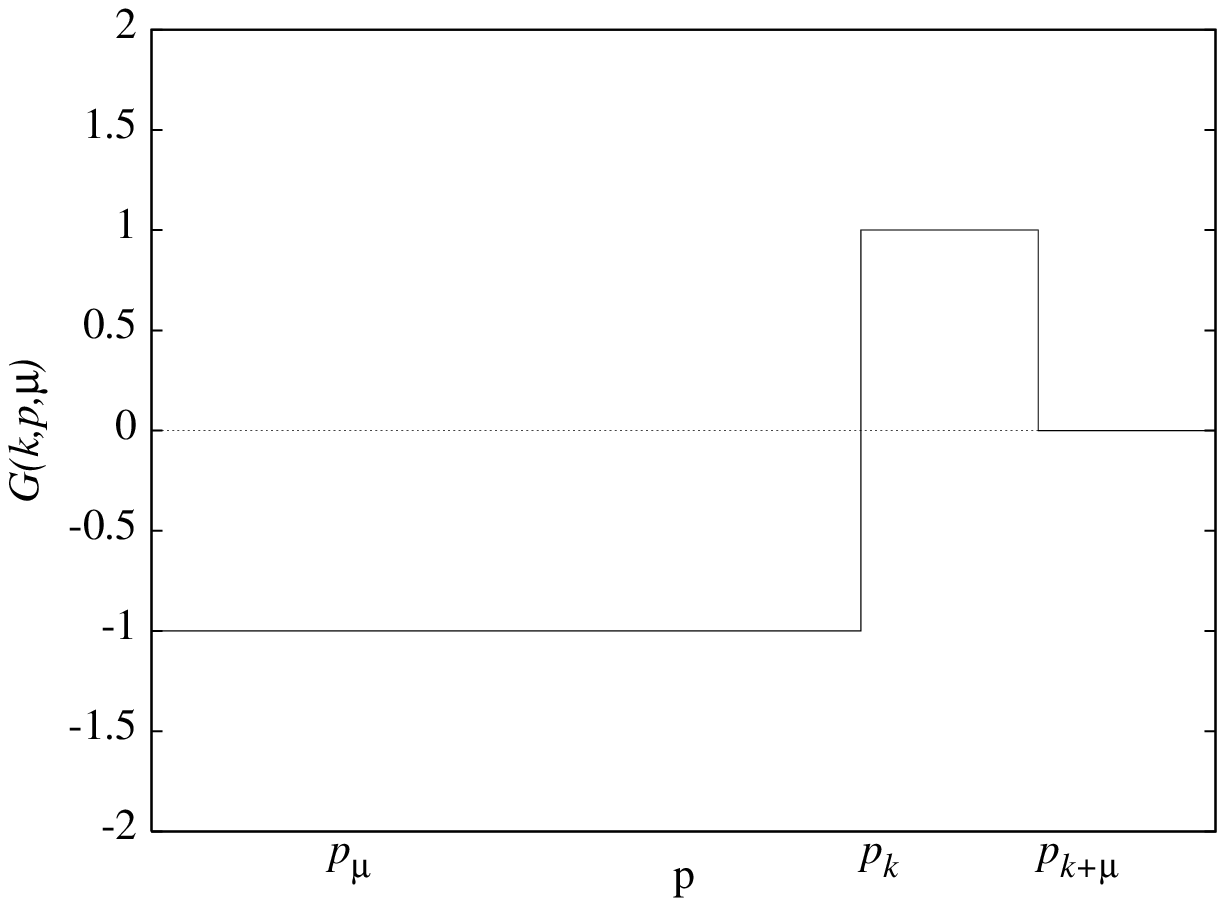,width=200pt}
\caption[]{${G}(k,p,\mu)$ vs. $p$ for a Fermi-degenerate plasma.}
\eef

\ni In this section we consider a degenerate fermion gas at zero temperature.
For degenerate fermions eq.(\ref{Y_f}) reduces to,
\ber
&& Y(\beta, k, E_p) + Y(\beta, k, -E_p) \nonumber \\
&=& 2 \, \epsilon(E_p) \, \coth \frac{\beta k_0}{2} \nonumber \\
&\times& \Bigg(\epsilon(E_p+ k_0) \big(\frac{1}{e^{\beta(E_p +k_0 - \mu)}+1} - \frac{1}{e^{\beta(E_p - \mu)}+1} \big) \nonumber \\
&+& \epsilon(E_p- k_0) \big(\frac{1}{e^{\beta(E_p -k_0 - \mu)}+1} - \frac{1}{e^{\beta(E_p - \mu)}+1} \big) \Bigg) \,.
\label{Y_deg}
\eer
Let us consider the special case of relativistic fermions (for example the relativistically degenerate electrons in 
white dwarfs or neutron stars) and very high energy photons such that $m^2 < k^2$ and $k_0 \pm K > 2 \mu$. 
With this assumption we have $P_{max}= K/2 + k_0/2$ and $P_{min} = k_0/2 - K/2$. Because of the high virtuality of the 
photon ($k^2 >> m^2$) we can assume that $k_0 >> K$. It is now easy to see that for a degenerate Fermi gas at $T=0$ the 
factor inside the bracket (let us denote it by $G(k, p, \mu)$ of eq.(\ref{Y_deg}) is given by fig.2, where we have 
used the following definitions:
\ber
p_{\mu} &=& \sqrt{\mu^2 - m^2}\,, \\
p_{k} &=& \sqrt{k_0^2 - m^2}\,, \\
p_{k+\mu} &=& \sqrt{(\mu + k_0)^2 - m^2}\,.
\eer
It is evident that in the interval [$P_{\rm min}, P_{\rm max}$], for our choice of the kinematics, $G$ is entirely
negative. Therefore the field dependent part of the thermal polarisation tensor takes the form: 
\ber
\Pi^{\rm abs}_{\mu\nu}(k, \beta)
&=&  \, \frac{\pi^2}{2} e^3 {\cal B} \varepsilon_{\mu\nu\alpha_\parallel\beta} k^\beta
\, \left(\partial_{k^{\widetilde\alpha_\parallel}} \, \frac{1}{K}\right) \nonumber \\
&\times& \int_{P_{\rm min}}^{P_{\rm max}} \frac{P dP}{2\pi} \frac{1}{E_p} \nonumber \\
&+& \frac{\pi^2}{2} e^3 {\cal B} \varepsilon_{\mu\nu\alpha_\parallel\beta} \frac{k^\beta}{K}
\left[\frac{P}{E_p} \right]_{P_{\rm max}}\partial_{k^{\widetilde\alpha_\parallel}} P_{\rm max} \nonumber \\
&-& \frac{\pi^2}{2} e^3 {\cal B} \varepsilon_{\mu\nu\alpha_\parallel\beta} \frac{k^\beta}{K}
\left[\frac{P}{E_p} \right]_{P_{\rm min}} \partial_{k^{\widetilde\alpha_\parallel}} P_{\rm min} \nonumber \\ 
&\simeq& \frac{\pi^2}{2} e^3 {\cal B} \varepsilon_{\mu\nu\alpha_\parallel\beta} k^\beta
\, \left(\partial_{k^{\widetilde\alpha_\parallel}} \, \frac{1}{K}\right) \nonumber \\
&\times& m \left[\f{1}{1+ \left( \f{k_0+K}{2m}\right)^2} -\f{1}{1+ \left( \f{k_0 - K}{2m}\right)^2} \right]
\label{e_pideg}
\eer
To derive the above expression we have assumed a very high energy photon i.e, $ P^2_{\rm max} + m^2 \sim P^2_{\rm min}+m^2$
in the limit of $k_0 >> K$. Further assuming the photon to be moving in a direction parallel to the direction of the magnetic 
field we obtain,
\ber
\Pi^{\rm abs}_{\mu\nu}(k,\beta) \sim 4 \pi^2 e^3 {\cal B} \varepsilon_{\mu\nu 0 3} \frac{m^3}{k^3_0}\,.  
\label{app-pol}
\eer
\ni{\bf Dispersive Part of $\Pi_{\mu \nu}$} -- Here we evaluate field dependent part of the real part of the vacuum 
polarisation tensor for a degenerate fermion gas. Our starting point is eq.(5.21) in \cite{GKP1}. For a weak magnetic 
field it can be shown that:
\ber
\Pi^{\rm disp}_{\mu\nu}(k,\beta) &=& \frac{\pi^2}{2} e^3 {\cal B} \varepsilon_{\mu\nu\alpha_\parallel\beta} k^\beta 
\int \frac{d^4p}{(2 \pi)^4} \, \epsilon(p_0) \, f_{+}(p) \nonumber \\
&\times& \int^{\infty}_{\infty} e^{is(p^2-m^2) -\epsilon |s|} \int^{\infty}_{0} e^{is'(p'^2-m^2) -\epsilon s'} \nonumber \\
&\times& \Bigg[\left(p^{\twidle\alpha_{\parallel}}(s+s') - 2 p^{\twidle\alpha_{\parallel}} s 
+ 2p^{\twidle\alpha_{\parallel}} \f{s^2}{s+s} \right) \nonumber \\
&+& k^{\twidle\alpha_{\parallel}} \left((s+s') -2s + \f{s^2}{(s+s')} \right) \Bigg].
\label{realpol}
\eer
Because of the kinematics and the condition that the photon propagates
along the z direction, the term proportional
to $k^{\twidle \alpha_{\parallel}}$  would vanish. Now the non-vanishing piece
comes from the term proportional to $p^{\twidle \alpha_{\parallel}}$.
Using arguments as has been used previously
in \cite{GKP1}, we can write, 
\ber
I = i \f{\partial}{\partial(m^2)} J_{0} - 2J_{1} +i \int {\rm d(m^2)}J_2
\eer
With the difference that contrary to~\cite{GKP1} here, powers of $s$ appear in the numerator instead of $s'$ in the 
second and third term. The $J$~s are defined as,
\ber
J_n &=& \int \frac{d^4p}{(2 \pi)^4} \epsilon(p_0) f_{+}(p) \int^{\infty}_{\infty} e^{is(p^2-m^2) -\epsilon |s|} \nonumber \\
&\times& \int^{\infty}_{0} e^{is'(p'^2-m^2) -\epsilon s'} s^n \,.
\label{jn}
\eer
Out of the three terms in eq.(\ref{realpol}) the last two terms do not contribute. Therefore, we are left with only 
the first term. The $s$ and $s'$ integration in $J_0$ leaves us with,
\ber
I &=& \f{\partial}{\partial(m^2)} \int \frac{d^4p}{(2 \pi)^4 } 
\epsilon(p_0)f_{+}(|p_0|) p^{\widetilde{\alpha}_{\parallel}} \f{\delta(p^2 - m^2)}{(p'^2 - m^2)}, \nonumber \\
&\simeq& \f{\partial}{\partial(m^2)} \int \frac{d^3p}{(2 \pi)^4 } f_{+}(|p_0|) \nonumber \\
&& \times \left(\f{1}{(k^2+ 2E_p k_0)} - \f{1}{(k^2-2E_p k_0)}\right) \,
\label{I}
\eer 
with $E_p = \sqrt{\vec{p^2}+ m^2}$. In the last step we have neglected terms proportional to $\vec{P} \cdot \vec{k}$ 
in the denominator, since $P^2_F,~|K|^2$ and $m^2$ are negligible compared to $k^2 \sim k^2_0$. The dominant behavior 
of eq.(\ref{realpol}) is,
\beq
I = - \f{P^2_F}{2 k^3_0}\,.
\label{I1}
\eeq  
So the real part of the of the polarisation tensor, in a weak magnetic field and degenerate fermion gas, comes out to be,
\ber
\Pi^{\rm disp}_{\mu\nu}(k, \beta)
=  -4i e^3 {\cal{B}}\,\varepsilon_{\mu\nu 3 0} \frac{P^2_F}{2 k^2_0}.
\eer
Now we can find out the lifetime of a photon in the plasma using eq.(\ref{disp2}). Note that for a degenerate plasma 
$\omega_0 \sim P_F$ and hence this is more dominant than the contribution from $\Pi^{abs}$. Therefore, we neglect $\Pi^{abs}$
while estimating photon damping time. With a little bit of algebra one can show that
\ber
Im (k_0) \sim 4 \pi^2 e^3 {\cal{B}}\f{m^3}{ \omega_0 k^3_0}\,, 
\label{damping}
\eer
the inverse of which is the life-time of the photon.
%%%%%%%%%%%%%%%%%%%%%%%%%%%%%%%%%%%%%%%%%%%%%%%%%%%%%%%%%%%%%%%%%%%%%%%%%
\section{conclusion}
%%%%%%%%%%%%%%%%%%%%%%%%%%%%%%%%%%%%%%%%%%%%%%%%%%%%%%%%%%%%%%%%%%%%%%%%%
\ni We have calculated the absorptive part of the polarisation tensor in a magnetised medium to 1-loop order.
Our most general result is odd in powers of $e{\cal B}$ summed to all orders. In that we have retained terms
up-to the linear order in $e{\cal B}$ and estimated the damping rate of a photon propagating along the magnetic
field. We have specialised to a degenerate fermionic system where the kinematics of the process under consideration 
is $k^2 > m^2$ and $k_0 \pm K >> 2 \mu$ with $T \sim 0$. In this domain we see that the imaginary part of 
$\omega$ has a positive sign which signifies that the propagating photon gets damped by creation of fermion anti-fermion 
pairs. If we take the inverse, it would give us the time scale of damping. Since $e{\cal B}/m^2$ is small the damping 
or attenuation time would be rather large. 
%%%%%%%%%%%%%%%%%%%%%%%%%%%%%%%%%%%%%%%%%%%%%%%%%%%%%%%%%%%%%%
\section*{Acknowledgment}
%%%%%%%%%%%%%%%%%%%%%%%%%%%%%%%%%%%%%%%%%%%%%%%%%%%%%%%%%%%%%
\ni We thank Palash B. Pal for many illuminating discussions and detailed comments on the manuscript. AKG acknowledges 
Ashok Das for useful comments and SK thanks Rajaram Nityananda \& S. Shankaranarayanan for helpful discussions. 
%%%%%%%%%%%%%%%%%%%%%%%%%%%%%%%%%%%%%%%%%%%%%%%%%%%%%%%%%%%%
\appendix \section*{}
%%%%%%%%%%%%%%%%%%%%%%%%%%%%%%%%%%%%%%%%%%%%%%%%%%%%%%%%%%%
\ni {\bf Evaluation of $I( p,p')$} - Recall that $I(p,p')$ is given by,
\ber
I(p,p') &=& \int_{-\infty}^\infty ds \, e^{\Phi(p,s)} \int_{-\infty}^\infty ds' \, e^{\Phi(p',s')} \, \frac{s'}{s'+s} \nonumber \\
&=& \frac{1}{2} \, \int_{-\infty}^\infty ds \, e^{\Phi(p,s)} \int_{-\infty}^\infty ds' \, e^{\Phi(p',s')} 
\, \left[1 + \frac{s'-s}{s'+s}\right] \nonumber \\ 
&=& 2 \pi^2 \delta(p^2 - m^2) \, \delta(p'^2 - m^2) + \frac{1}{2} \, I'(p,p')
\eer
where we have defined :
\ber
I'(p,p') &=&  \int_{-\infty}^\infty ds \, e^{\Phi(p,s)} \int_{-\infty}^\infty ds' \, e^{\Phi(p',s')} 
\, \left[\frac{s'-s}{s'+s}\right] \nonumber \\
&=& \int_{-\infty}^\infty \frac{dt}{t} \, e^{it(p^2 + p.k + k^2/2 - m^2-i \epsilon_1)} \nonumber \\
&\times& \, \int_{-\infty}^\infty dt' \, t' \, e^{it'( p.k + k^2/2 -i\epsilon_2)}
\label{I'}
\eer
where $t=s+s'$ and $t'=s'-s$. Now,
\ber
&& \int_{-\infty}^\infty dt' \, t' \, e^{it'( p.k + k^2/2 -i\epsilon_2)} \nonumber \\
&=& \frac{1}{2i\pi} \,\left\{\frac{1}{( p.k + k^2/2 +i\epsilon_2)^2} -\frac{1}{( p.k + k^2/2 -i\epsilon_2)^2}\right\} \nonumber \\
%&=& \frac{1}{2i\pi} \, \frac{4 i \epsilon (p.k + k^2/2)}{\left\{(p.k + k^2/2)^2 + \epsilon_2^2\right\}^2} \nonumber \\
%&=& \frac{1}{\pi} \, \frac{2 \epsilon (p.k + k^2/2)}{(p.k + k^2/2)^2 + \epsilon_2^2} \, 
%\, \frac{1}{2 i \epsilon} \left[ \frac{1}{p.k + k^2/2 - i\epsilon_2} - \frac{1}{p.k + k^2/2 + i\epsilon_2} \right]\nonumber \\
&=& \frac{2 (p.k + k^2/2)}{(p.k + k^2/2)^2 + \epsilon_2^2} \, \delta(p.k + k^2/2)
\eer
where we have used the following identity :
\beq
\frac{1}{a \pm i \epsilon} = {\cal P}(a) \mp i \pi \delta(a) \,, \, \mbox{$\cal P$ being the principal value}.
\eeq
Therefore we have, 
\ber
I'(p,p') &=& \delta(p.k + k^2/2) \, \frac{2(p.k + k^2/2)}{(p.k + k^2/2)^2 + (\epsilon_2)^2} \nonumber \\
&\times& \int_{-\infty}^\infty \frac{dt}{t} \, e^{it(p^2 + p.k + k^2/2 - m^2-i \epsilon_1)}
\label{B2}
\eer
Since the numerator and the argument of the delta function are the same, $I'(p,p')$ vanishes upon $p$-integration, provided
we take the limit $\epsilon_2 \rightarrow 0^+$ later. Therefore, finally we are left with :
\beq
I(p,p') = 2 \pi^2 \delta(p^2 - m^2) \, \delta(p'^2 - m^2) \,.
\eeq

%%%%%%%%%%%%%%%%%%%%%%%%%%%%%%%%%%%%%%%%%%%%%%%%
\beb
\bi{melr} E. Parker, {\em Cosmic Magnetic Fields}, (Oxford University Press, 1979);
          Ya.~B. Zeldovitch, A.~A. Ruzmaikin, D.~D. Sokoloff, {\em Magnetic Fields in Astrophysics}, (Gordon \& Breach, 1983);
          J.~G. Kirk, D.~B. Melrose, E.~R. Priest, A.~O. Benzo, T.~J.~L. Courvoisier, {\em Plasma Astrophysics : Saas-Fee Advanced
          Course 24}, (Springer-Verlag, 1984);
          L. Mestel, {\em Stellar Magnetism}, (Oxford University Press, 1999)
\bi{kouv} C. Kouveliotou, T. Strohmayer, K. Hurley, J. van Paradijs, M.~H. Finger, S. Dieters,
P. Woods, C. Thompson, R.~T. Duncan, Astrophys. J. 581 (1999) L103 
\bi{chan} G. Chanmugam, Ann. Rev. Astron. \& Astrophys., 30 (1992) 143
\bi{GKP1} A.~K. Ganguly, S. Konar and P.~B. Pal, Phys. Rev. D60 (1999) 105014
\bi{matsu} T.~Matsubara, Prog. Theor. Phys. 14 (1955) 351
%\bi{NPeps} J.~F. Nieves, P.~B. Pal, Phys. Rev. D39 (1989) 652; Erratum ibid D40 (1989) 2148
\bi{larry} L.~D. Mclerran, T. Toimela, Phys. Rev. D49 (1985) 1047
\bi{kapusta} J. Kapusta, {\em Finite Temperature Field Theory }, (Cambridge University Press, 1989)
\bi{lebellac} M. Le Bellac, {\em Thermal Field theory}, (Cambridge University Press, 1996)
\bi{adas} A. Das, {\em Finite Temperature Field Theory}, (World Scientific, 1997)
\bi{kobes} R. Kobes, Phys. Rev. D42 (1990) 562; {\it ibid.} D43 (1991) 1269
\bi{Gross} D.~J. Gross, R.~D. Pisarski and L.~G. Yaffe, Rev. Mod. Phys. 53 (1981) 43
\bi{jose} J.~F. Nieves, Phys. Rev. D42 (1990) 4123
\bi{land} N.~P. Landsman \& Ch.~G. van Weert, Phys. Rep. 145 (1987) 141; 
               P. Aurenche, T. Becherrawy, Nuc. Phys. B379 (1992) 259; 
               R. Kobes, Phys. Rev. D43 (1991) 1269; 
               N. Ashida, H. Nakkagawa, A. Niegawa, H. Yokota, Phys. Rev. D45 (1992) 2066  
\bi{RojSha79} H. P\'erez Rojas, A.~E. Shabad, Ann. Phys. 121 (1979) 432
%\bi{epsmu} M.~B. Kislinger, P.~D. Morley, Phys. Rev. D13 (1976) 2765 and 2771; 
                H.~A. Weldon, Phys. Rev. D26 (1982) 1394
\bi{Schwing} J. Schwinger, Phys. Rev. 82 (1951) 664
\bi{Tsai} W.~Y. Tsai, Phys. Rev. D10 (1974) 1342 and 2699
\bi{Dittrich} W. Dittrich, Phys. Rev. D19 (1979) 2385
\bi{Elmf} P. Elmfors, D. Grasso, G. Raffelt, Nucl. Phys. B479 (1996) 3
\eeb
%%%%%%%%%%%%%%%%%%%%%%%%%%%%%%%%%%%%%%%%%%%%%%%%

\end{document}